\documentclass[twocolumn]{jpsj2}

\def\runauthor{Takehiro \textsc{Saito} {\it et al.}}

\title{High-field Magnetic Torque Measurements in the Spin Gap System (CH$_3$)$_2$CHNH$_3$CuCl$_3$}

\author{Takehiro \textsc{Saito}$^1$, Takahiko \textsc{Sasaki}$^2$, Takao \textsc{Suzuki}$^{3,1}$, Akira \textsc{Oosawa}$^1$, Takayuki \textsc{Goto}$^{1,2,}$\thanks{E-mail address: gotoo-t@sophia.ac.jp}, Satoshi \textsc{Awaji}$^2$, Kazuo \textsc{Watanabe}$^2$ and Norio \textsc{Kobayashi}$^2$}

\inst{$^1$Department of Physics, Sophia University, 7-1 Kioi-cho, Chiyoda-ku, Tokyo 102-8554, Japan\\
$^2$Institute for Materials Research, Tohoku University, Sendai 980-8577, Japan\\
$^3$Advanced Meson Science Laboratory, RIKEN, 2-1 Hirosawa, Wako, Saitama 351-0198, Japan\\}

\recdate{\today}

\abst{High-field magnetic torque measurements were carried out in the spin gap system (CH$_3$)$_2$CHNH$_3$CuCl$_3$. It was observed that the magnitude of the magnetic torque $\tau$ is almost zero until the critical field and then increases rapidly, which indicates the existence of the magnetic quantum phase transition from the spin-gap phase to the field-induced magnetic ordered phase. From the temperature dependence of the magnetic torque $\tau$, the cusp-like minimum indicative of the Bose-Einstein condensation (BEC) of magnons was observed for $H \perp$ C-plane, while the kink-anomaly was observed for $H \perp$ A-plane at the field-induced transition temperature. It was found that this different behavior between $H \perp$ A-plane and  $H \perp$ C-plane can be interpreted as the breaking of the magnon BEC picture due to the rotational symmetry breaking. The additional anomaly was also observed in the field-induced magnetic ordered phase at the intermediate angle of the applied magnetic field in the A-plane and discussed in terms of the spin-flop transition, the spin-reorientation transition and the spin-lattice correlations. In order to investigate the spin-lattice correlations in (CH$_3$)$_2$CHNH$_3$CuCl$_3$, we performed the magnetostriction measurements and observed that the magnetostriction appears by vanishing of the spin gap.}

\kword{(CH$_3$)$_2$CHNH$_3$CuCl$_3$, spin gap, field-induced magnetic ordering, Bose-Einstein condensation of magnons, anisotropy, high-field magnetic torque measurement, magnetostriction measurement}

\begin{document}
\sloppy
\maketitle

\section{Introduction}

The field-induced magnetic ordering of spin gap systems has attracted much attention from the viewpoint of the quantum phase transition between the nonmagnetic disordered phase with spin gap and the magnetic ordered phase which can be interpreted as the Bose-Einstein condensation (BEC) of magnons \cite{Nikuni}. \par
The title compound (CH$_3$)$_2$CHNH$_3$CuCl$_3$ (abbreviated as IPACuCl$_3$) is the spin gap system. By means of the magnetic susceptibility measurements, the magnitude of the excitation gap $\Delta$ was estimated as $17.1 \sim 18.1$ K in IPACuCl$_3$ \cite{Manakasus}. From the viewpoint of the crystal structure, the origin of the spin gap was expected to be the $S=\frac{1}{2}$ ferromagnetic-antiferromagnetic alternating chain along the $c$-axis \cite{Manakasus}. However, quite recently, it was suggested that this system should be characterized as the spin ladder along the $a$-axis with the strongly coupled ferromagnetic rungs, namely the antiferromagnetic chain with effective $S=1$, and the excitation gap $\Delta$ was re-estimated as 13.6 K by means of the neutron inelastic scattering experiments \cite{Masuda}. By means of the high-field magnetization processes, it was observed that the magnetization is almost zero until the critical field normalized by the $g$-factor $H_c=10.4$ T due to the gapped singlet ground state, and then increases with no anomalies and saturates at around $H_s=44$ T with increasing magnetic field \cite{Manakamag}. \par
When a magnetic field is applied in spin gap systems, the excited triplet states split due to the Zeeman interaction so that the energy gap $\Delta$ is suppressed. Consequently, the energy gap $\Delta$ closes at the field $H_{\rm g}$ corresponding to the energy gap. For $H>H_{\rm g}$, the ground state becomes magnetic and the system can undergo the magnetic ordering with the help of the three-dimensional interactions. This is the mechanism of the field-induced magnetic ordering of spin gap systems. Such field-induced magnetic ordering has been actually observed in IPACuCl$_3$ by means of the specific heat measurements \cite{Manakaheat}. The normalized critical field by the $g$-factor corresponding to the transition field at $T=0$ K was estimated as $H_{\rm g}=10.1$ T by extrapolating the obtained phase boundary. The $H_{\rm g}$ corresponds to the $H_c$ estimated in the high-field magnetization processes \cite{Manakamag}. \par
As mentioned above, the field-induced magnetic ordering can be interpreted as the BEC of magnons. However, the magnon BEC picture is only applied to spin gap systems in which there is rotational invariance for the direction of the applied magnetic field in order to conserve the number of magnons \cite{Nikuni}. This means that the existence of the anisotropy affects the property of the field-induced magnetic ordering, namely whether or not the ordering can be interpreted as the magnon BEC. \par 
Recently, it was found that the excited triplet states already split at zero field due to the easy-axis-type fictitious single ion anisotropy $D^*$ caused by the dipole-dipole interaction and the anisotropic exchange interaction between ferromagnetically coupled two $S=\frac{1}{2}$ spins in IPACuCl$_3$ \cite{ManakaESR1, ManakaESR2}. Hence we can expect that the anisotropy affects the field-induced magnetic ordering in IPACuCl$_3$. \par
In this paper, we report the results of the investigation of magnetic properties at high magnetic fields in IPACuCl$_3$ by means of the magnetic torque measurement, which is quite sensitive for magnetic properties related to the anisotropies so that this measurement has been carried out for the investigations of magnetic phase transitions, such as the spin-flop transition, in many magnetic systems. We also measured the magnetostriction for high magnetic fields in order to discuss the spin-lattice correlations in IPACuCl$_3$. \par

\section{Experimental Details}

The single crystals of IPACuCl$_3$ were prepared by dissolving (CH$_3$)$_2$CHNH$_2$ $\cdot$ HCl and CuCl$_2$ in isopropyl alcohol \cite{Manakasus}. The solution was slowly evaporated at 30 $^{\circ}$C over two months. The crystals with three orthogonal surfaces were obtained. We call these three planes as A-, B- and C-planes. The definition of these planes is shown in ref. \citen{Manakasus}. It was reported that the space group of this system is triclinic $P \bar{1}$ \cite{Manakasus}. The relation between these planes and the crystal axes are determined as that the $c$- and $b^*$-axes are perpendicular to the A- and C-planes, respectively, and the $a$-axis make an angle of $\sim 10^{\circ}$ in the C-plane with the direction perpendicular to the B-plane \cite{Manaka3DESR, ManakaMasuda}. \par
\begin{figure}[t]
\begin{center}
\includegraphics[width=80mm]{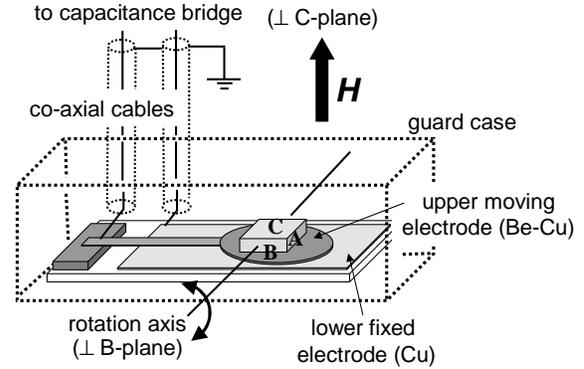}\\
\end{center}
\caption{Schematic picture of the unit with sample of IPACuCl$_3$ used for the magnetic torque measurements. In this configuration, the magnetic field is applied perpendicular to the C-plane and the magnetic torque projected to the direction perpendicular to the B-plane can only be detected. \label{Fig0}}
\end{figure}

The magnetic torque was measured using a cantilever beam torque meter \cite{Sasaki}. The capacitor consists of a circular moving Be-Cu electrode suspended by a narrow beam and a rectangular fixed Cu metal ground plate. They are surrounded by a small metallic case for the capacitive guard. A single crystal with 1 $\times$ 1 $\times$ 0.5 mm$^3$ was used for the magnetic torque measurements. The used single crystal was fixed on the moving electrode with a small amount of grease. The amplitude of the torque is obtained by measuring the capacitance with the use of a lock-in amplifier and a capacitance bridge. Figure \ref{Fig0} shows the schematic picture of the used unit with sample. The sample mounted on the moving electrode is only rotated about the rotation axis of moving electrode so that the magnetic torque $\tau$ parallel to the rotation axis is only detectable, namely the component of the magnetic torque perpendicular to the rotation axis cannot be observed. The measurements were carried out down to 0.4 K in magnetic fields up to 14.5 T using the 15 T-SM superconducting magnet with $^3$He refrigerator at High Field Laboratory for Superconducting Materials, Institute for Materials Research, Tohoku University. \par
The magnetostriction was measured by the strain gauge method using the relation of $\kappa \Delta L / L = \Delta R /R$ ($\kappa$ = 2.1, $L$ = 2 mm, $R$ = 118 $\Omega$). The gauge was glued onto a surface of the single crystal with 3 $\times$ 6 $\times$ 15 mm$^3$ by the grease. The field dependence measurements were carried out using the 12 T superconducting magnet settled in Department of Physics, Sophia University.

\section{Results and Discussion}

\subsection{Magnetic Torque Measurements}

\begin{fullfigure}[t]
\begin{minipage}{8.5cm}
\begin{center}
\includegraphics[width=80mm]{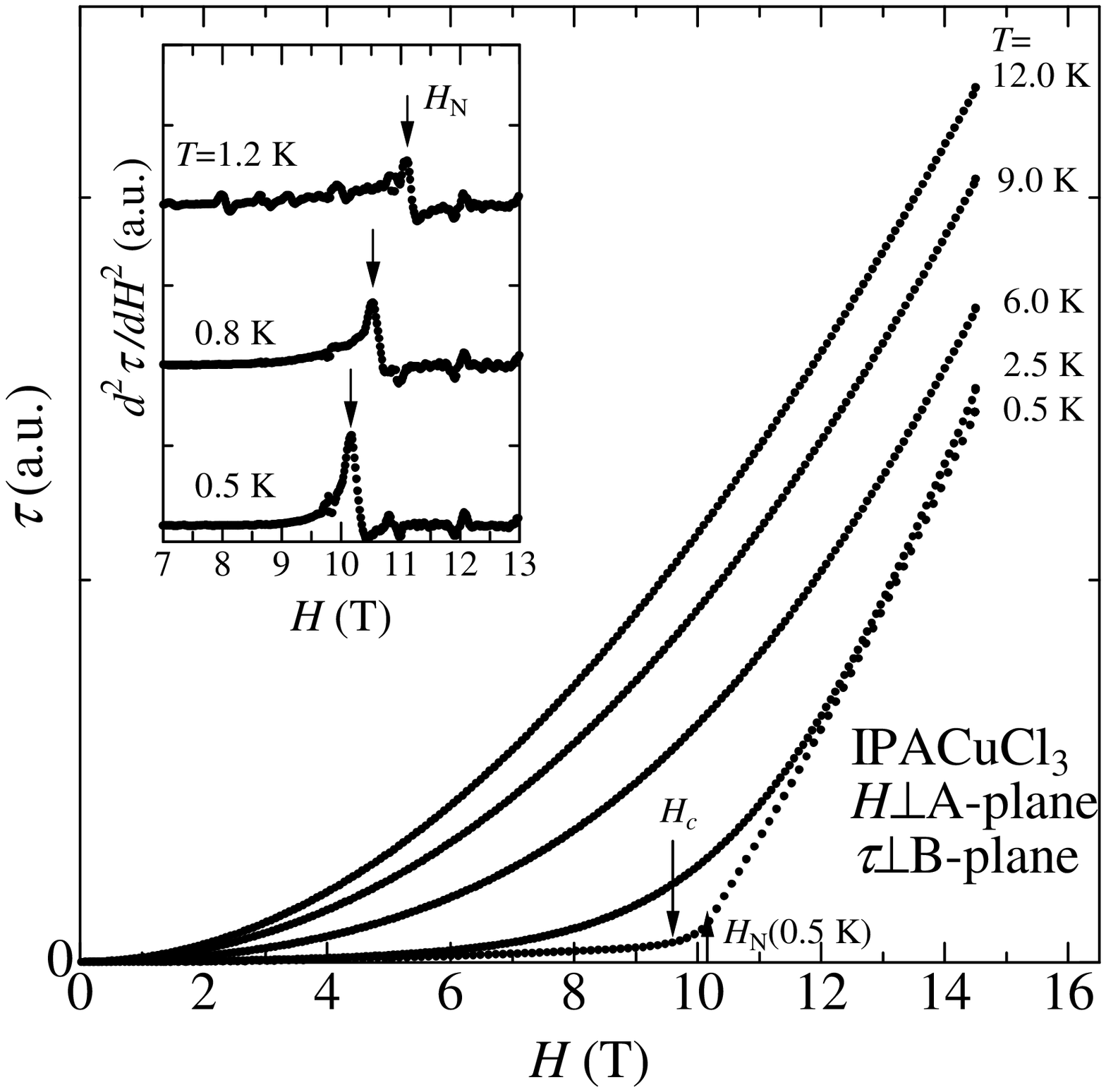}\\
(a)\\
\end{center}
\end{minipage}
\begin{minipage}{8.5cm}
\begin{center}
\includegraphics[width=80mm]{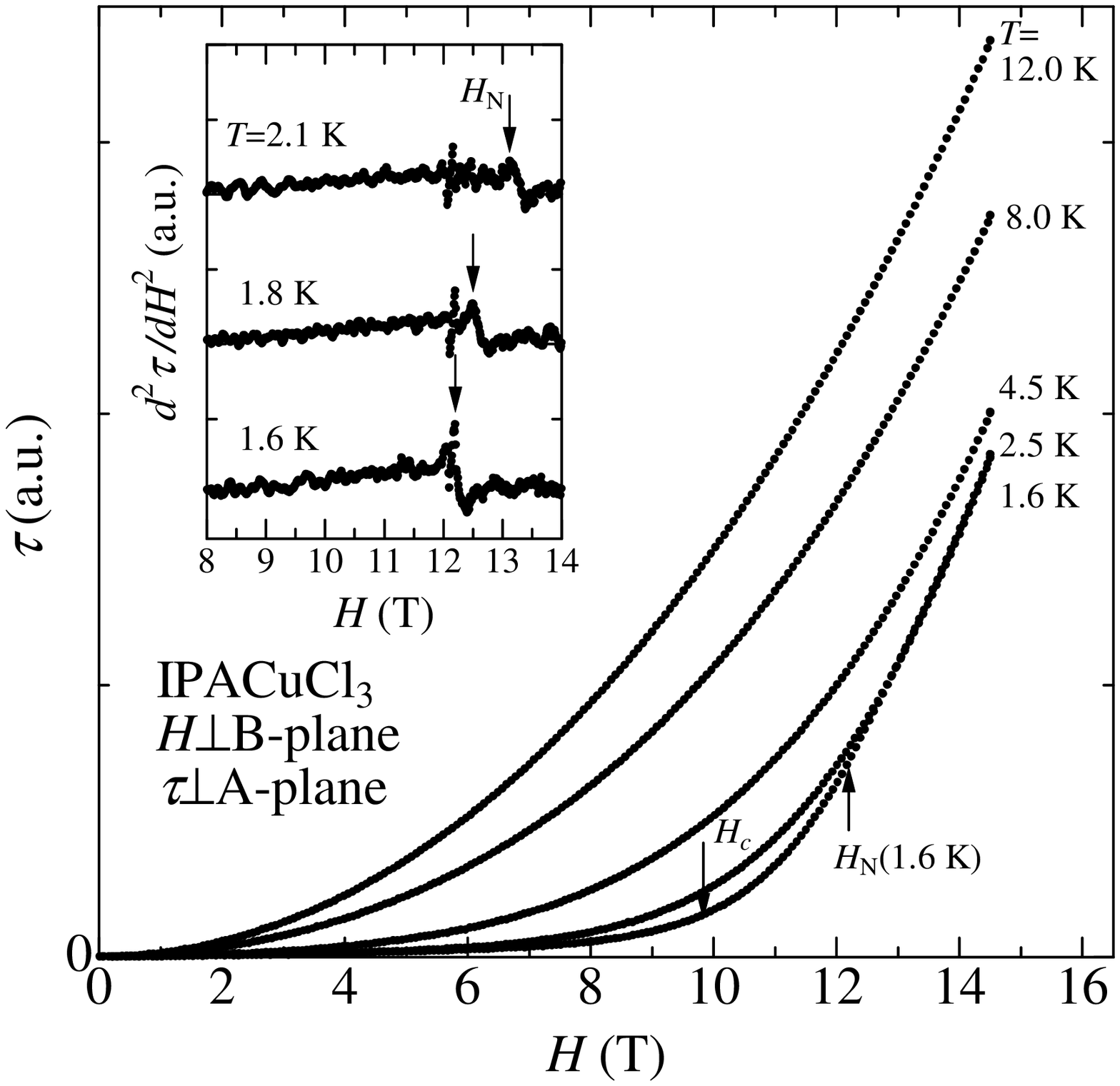}\\
(b)\\
\end{center}
\end{minipage}\\
\begin{minipage}{8.5cm}
\begin{center}
\includegraphics[width=80mm]{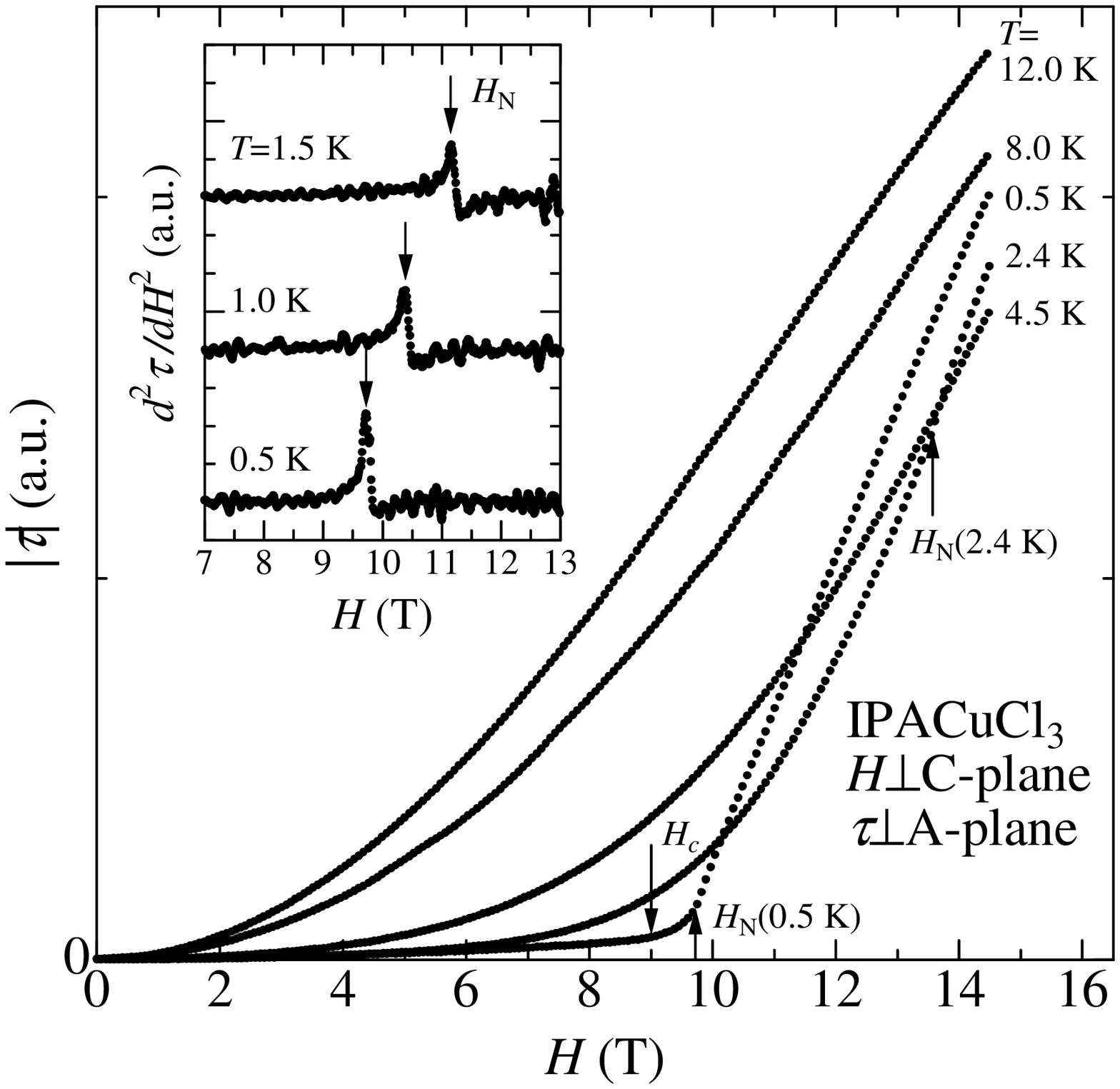}\\
(c)\\
\end{center}
\end{minipage}
\begin{minipage}{8.5cm}
\begin{center}
\includegraphics[width=80mm]{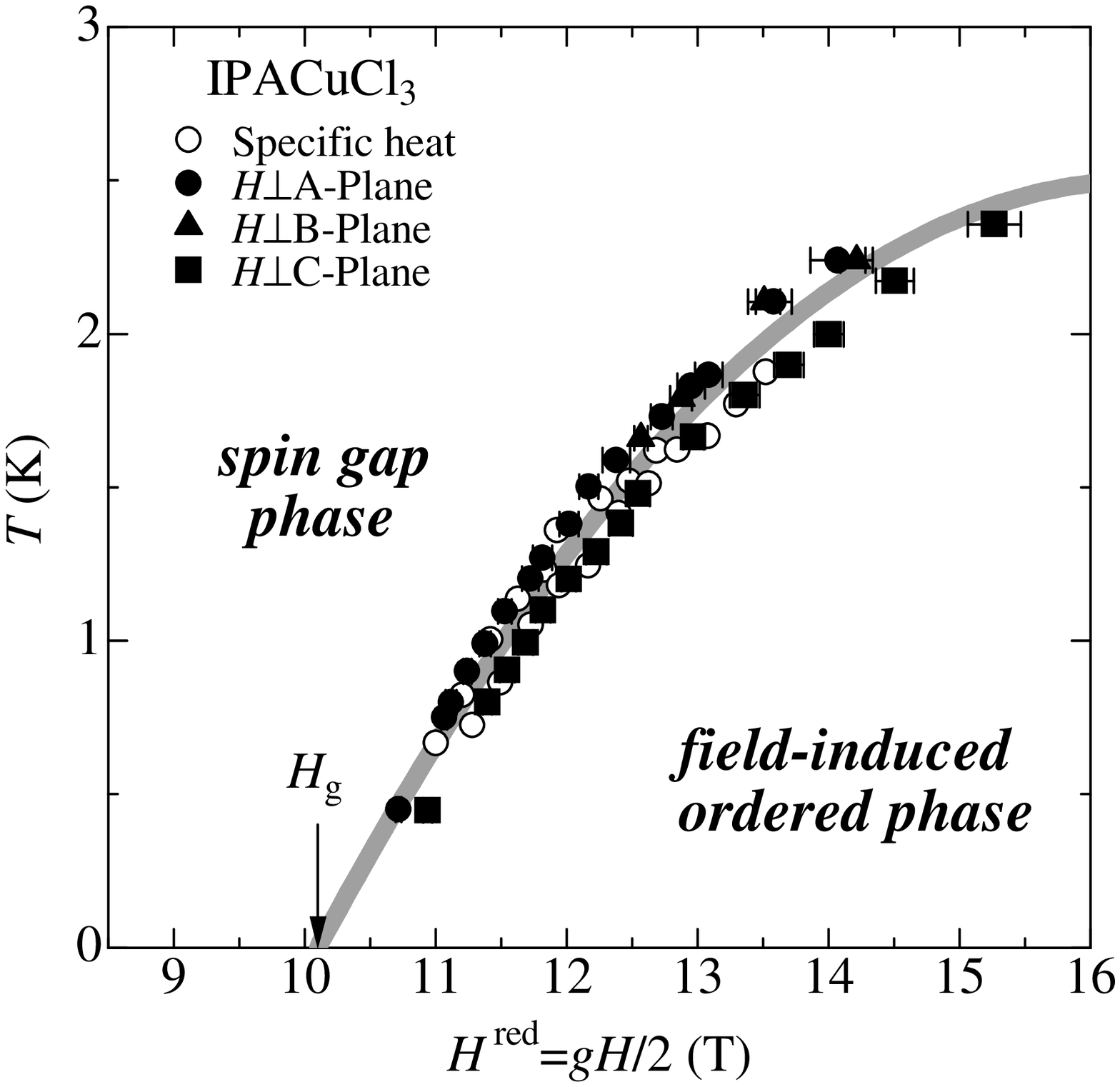}\\
(d)\\
\end{center}
\end{minipage}
\caption{Representative profiles of magnetic field dependence of the magnetic torque $\tau$ in IPACuCl$_3$ at various temperatures for $H \perp$ (a) A-plane, (b) B-plane, (c) C-plane, respectively. The detectable component of torque is also shown in each figures.  The absolute value is plotted for clearer comparison because the induced magnetic torque $\tau$ is negative for $H \perp$ C-plane. The insets show the second-field derivatives of magnetic torque $d^2 \tau/d H^2$ at various temperatures in IPACuCl$_3$. $H_c$ and $H_{\rm N} (T)$ denote the critical field corresponding to the spin gap \cite{Manakamag} and the transition field for the field-induced magnetic ordering determined from the field dependence of second-field derivatives of magnetic torque $d^2 \tau/d H^2$ (see in the text). (d) The obtained $H$-$T$ phase diagram of IPACuCl$_3$ where $H^{\rm red}$ and $H_{\rm g}$ denote the magnetic field normalized by the $g$-factor and the normalized critical field corresponding to the transition field at $T=0$ K, respectively \cite{ManakaESR1,Manakaheat}. The thick gray line denotes the phase boundary (guide for the eyes). The transition points obtained in the specific-heat measurements \cite{Manakaheat} are also plotted. \label{Fig1}}
\end{fullfigure}

Figure \ref{Fig1} (a), (b) and (c) show the representative profiles of magnetic field dependence of the magnetic torque $\tau$ in IPACuCl$_3$ at various temperatures for $H \perp$ A-, B- and C-planes, respectively. We observed the induced magnetic torque $\tau$ with positive for $H \perp$ A- and B-planes, while that with negative for $H \perp$ C-plane. Because the crystal structure of IPACuCl$_3$ is triclinic, the crystal axes and the principal axes of anisotropy tensor do not correspond to the three orthogonal surfaces of crystal A-, B- and C-planes so that it is difficult to obtain the information of these axes from the shape of the used crystal and the obtained data only for $H \perp$ A-, B- and C-planes. Hence, for simple comparison, the absolute value of the magnetic torque $\tau$ is plotted for $H \perp$ C-plane. As shown in Fig. \ref{Fig1} (a), (b) and (c), the magnitude of magnetic torque is almost zero until the critical field $H_c$, and then increases rapidly with increasing magnetic field. Also, the rapid increase is broadened so that the magnetic torque $\tau$ begins to evolve at the smaller critical field with increasing temperature. When the magnetic field $H$ is applied with the angle $\theta$ between $H$ and the principal axis of anisotropy $\xi$, the induced magnetic torque $\tau$ is expressed as 
\begin{eqnarray}
\label{B}
\tau &=& - \frac{\partial F}{\partial \theta} = - \frac{1}{2} \left( \chi_{\parallel \xi} - \chi_{\perp \xi} \right) H^2 \sin 2 \theta \nonumber \\
    &=& - \frac{1}{2} \left( M_{\parallel \xi} - M_{\perp \xi} \right) H \sin 2 \theta
\end{eqnarray}
\begin{fullfigure}[t]
\begin{minipage}{8.5cm}
\begin{center}
\includegraphics[width=80mm]{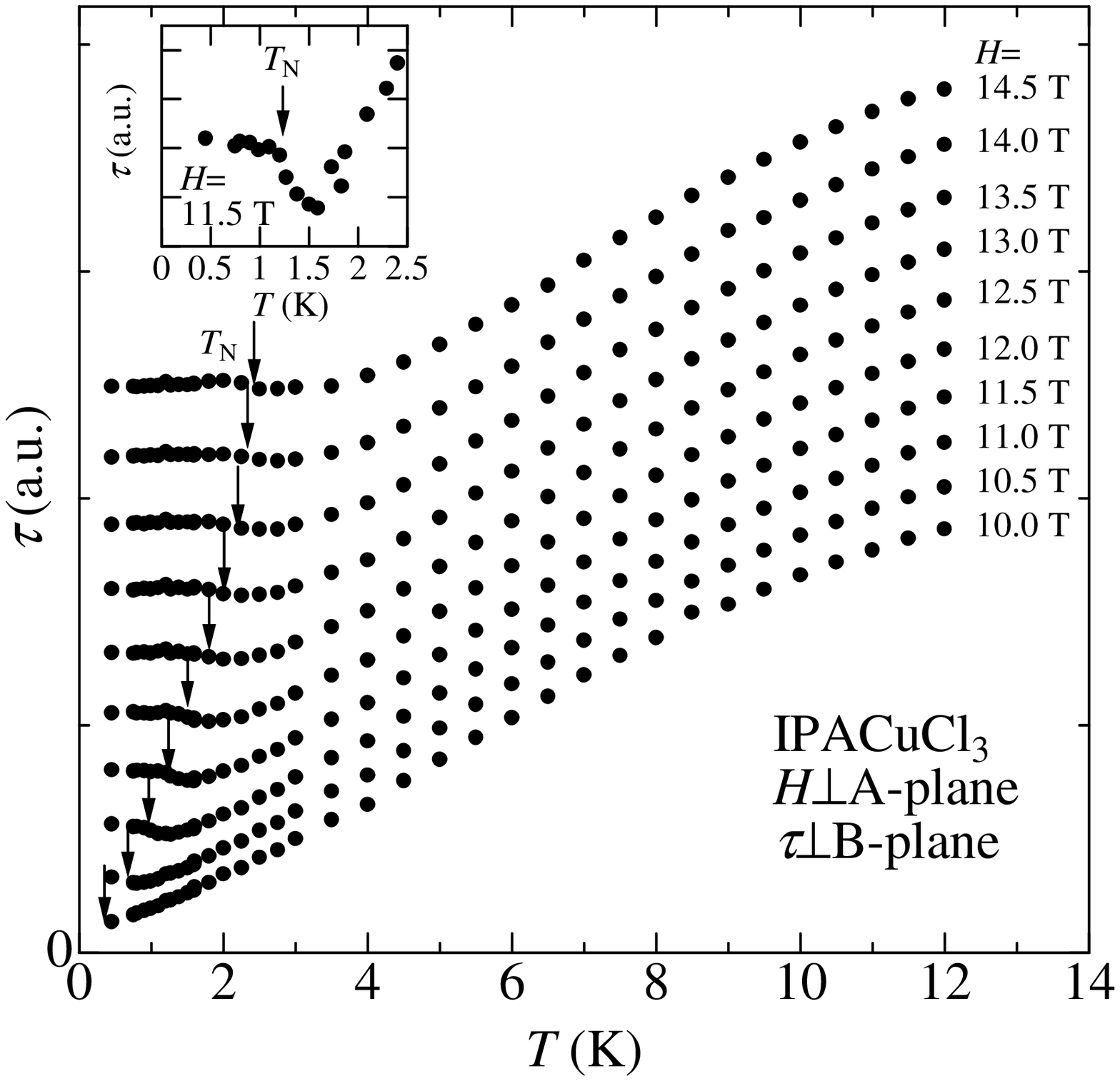}\\
(a)\\
\end{center}
\end{minipage}
\begin{minipage}{8.5cm}
\begin{center}
\includegraphics[width=80mm]{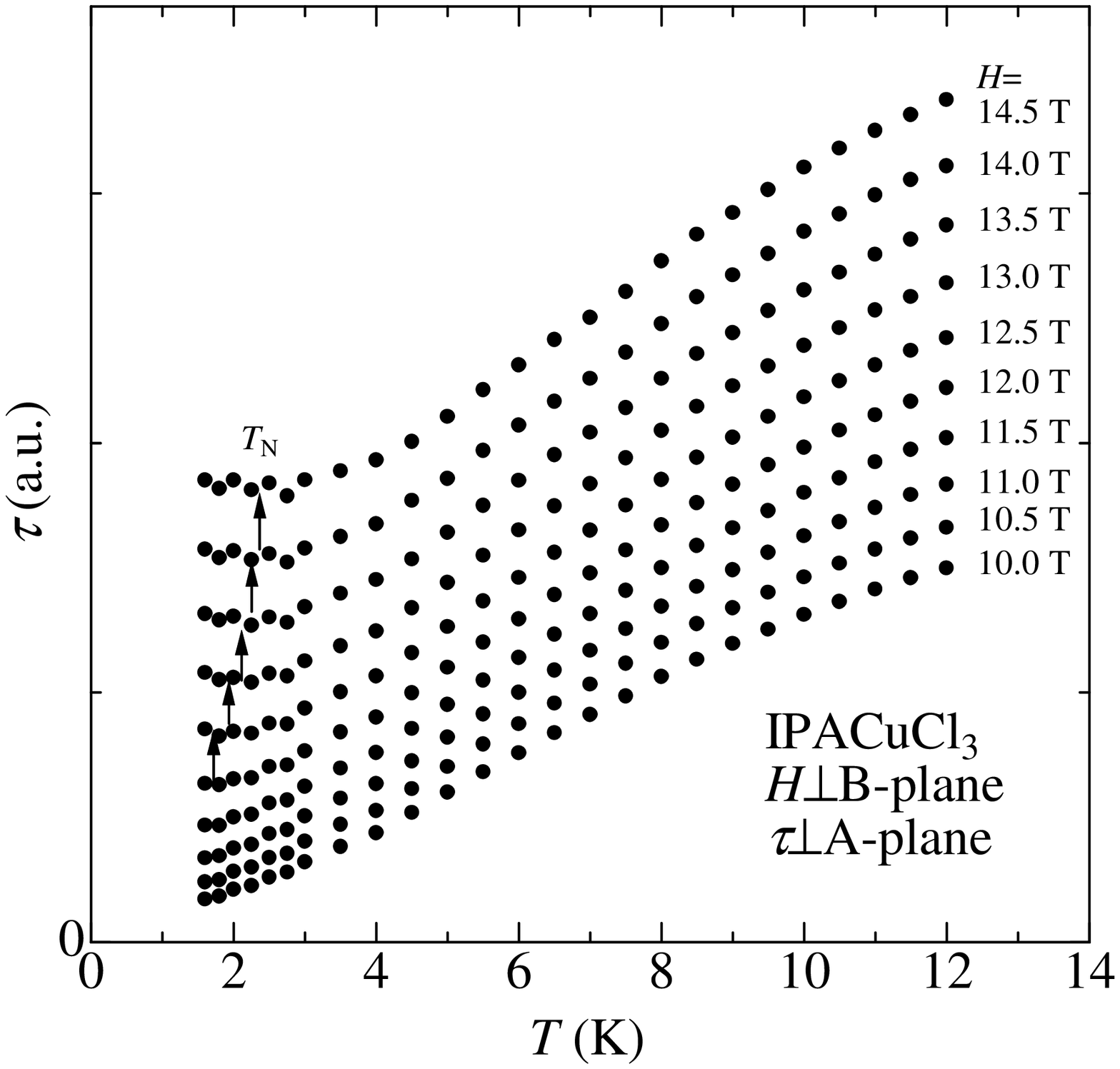}\\
(b)\\
\end{center}
\end{minipage}\\
\begin{center}
\includegraphics[width=80mm]{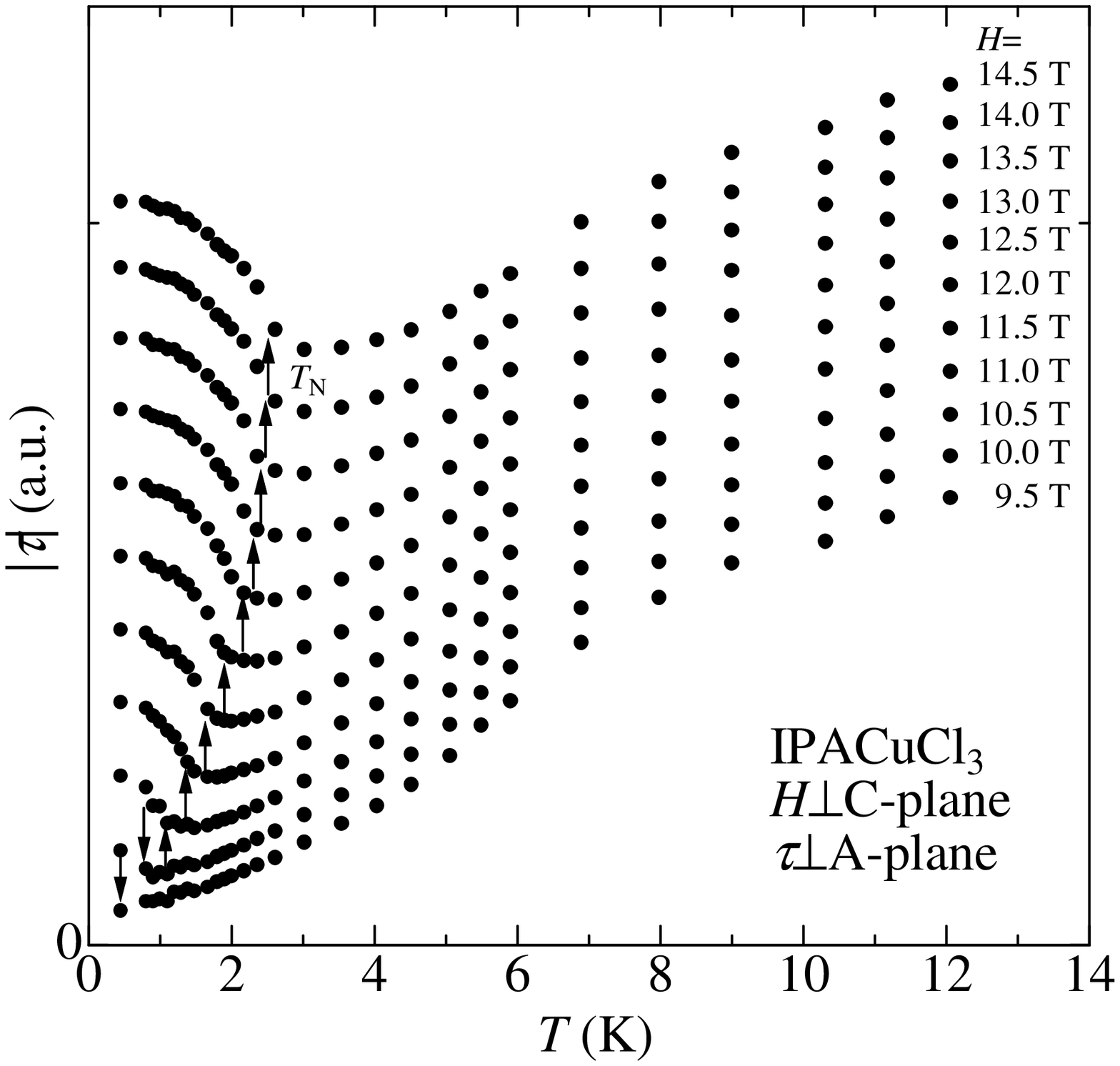}\\
(c)\\
\end{center}
\caption{Representative profiles of temperature dependence of the magnetic torque $\tau$ in IPACuCl$_3$ at various magnetic fields for $H \perp$ (a) A-plane, (b) B-plane, (c) C-plane, respectively. The absolute value is plotted for clearer comparison because the induced magnetic torque $\tau$ is negative for $H \perp$ C-plane. The inset of Fig. \ref{Fig2} (a) shows the enlarged profile of temperature dependence of the magnetic torque $\tau$ at $H=11.5$ T for $H \perp$ A-plane in IPACuCl$_3$. $T_{\rm N}$ denotes the transition temperature for the field-induced magnetic ordering determined from the phase boundary as shown in Fig. \ref{Fig1} (d). \label{Fig2}}
\end{fullfigure}
where $\chi_{\parallel \xi}$ and $M_{\parallel \xi}$ are the susceptibility and the magnetization for $H \parallel \xi$, and $\chi_{\perp \xi}$ and $M_{\perp \xi}$ are the susceptibility and the magnetization for $H \perp \xi$, respectively, and $F$ is the free energy. Note that the anisotropy energies are not considered in eq. (\ref{B}). We can expect from eq. (\ref{B}) that the magnetic torque $\tau$ is induced when the magnetization emerges because there is the anisotropy of the $g$-factors $g_{\perp A}$, $g_{\perp B}$ and $g_{\perp C}$, which have been already estimated as 2.11, 2.06 and 2.25, respectively, in IPACuCl$_3$ \cite{ManakaESR1}. Also, the fictitious single-ion anisotropy $D^*$ perpendicular to the B-plane exists \cite{ManakaESR1,ManakaESR2} and the principal axes of $g$ tensor correspond with neither the crystal axes nor the three orthogonal A-, B- and C-planes \cite{Manaka3DESR} so that it is quite difficult to determine the easy-axis of whole anisotropy in this triclinic system. Because the critical field $H_c$, at which the rapid increase of magnetic torque appears, corresponds to the critical field $H_c$ determined in the high-field magnetization processes \cite{Manakamag}, the rapid increase observed in the present experiments is due to the appearance of the magnetization by the vanishing of the spin gap. Because this system undergoes the field-induced magnetic ordering at low temperature above the critical field $H_c$ \cite{Manakaheat}, it can be expected that the magnetic torque $\tau$ shows the anomaly at the transition field $H_{\rm N}$. The insets of Fig. \ref{Fig1} show the second-field derivatives of magnetic torque $d^2 \tau/d H^2$ at various temperatures in IPACuCl$_3$. As shown in the insets of Fig. \ref{Fig1}, the sharp anomaly indicative of the phase transition can be clearly seen. Such anomaly has been also observed for the field-induced magnetic ordering in other spin gap systems \cite{Oosawa, Nikuni, Sebastian}. We assign the fields at which the second-field derivative of magnetic torque $d^2 \tau/d H^2$ shows the sharp peak to the transition field $H_{\rm N} (T)$. The obtained transition fields $H_{\rm N} (T)$ for $H \perp$ A-, B- and C-planes are plotted in the phase diagram, as shown in Fig. \ref{Fig1} (d). We can see that the obtained transition fields $H_{\rm N} (T)$ correspond well to the transition points of the field-induced magnetic ordering obtained in the specific heat measurements \cite{Manakaheat} when $H_{\rm N} (T)$ is normalized by the $g$-factor. From these results, we conclude that the field-induced magnetic ordering in IPACuCl$_3$ can be detected by means of the present magnetic torque measurements. \par
As shown in Fig. \ref{Fig1} (a), (b) and (c), it can be seen in high magnetic fields well above the critical field $H_c$ that the magnetic torque $\tau$ decreases with decreasing temperature and becomes independent on the temperature for both $H \perp$ A- and B-planes, while, shows the minimum at $T=4.5$ K after decreasing, and then re-increases with decreasing temperature for $H \perp$ C-plane. To show these behaviors more clearly, we re-plot the obtained field dependence of the magnetic torque $\tau$ at various temperatures as the temperature dependence of the magnetic torque $\tau$ at various magnetic fields, as shown in Fig. \ref{Fig2}. First, the magnetic torque $\tau$ decreases with decreasing temperature due to the development of the antiferromagnetic correlations associated with the formation of the spin gap for all directions. After that, the magnetic torque $\tau$ shows the cusp-like minimum, and then re-increases for $H \perp$ C-plane. This behavior is typical of the BEC of magnons, in which the magnetization shows the cusp-like minimum at the transition temperature in the temperature dependence, as observed in some spin gap systems \cite{Oosawa, Nikuni, OosawaK, Waki}. The arrows in Fig. \ref{Fig2} (c) denote the transition temperature obtained from the phase diagram determined above as shown in Fig. \ref{Fig1} (d). It can be seen that the temperature where the magnetic torque $\tau$ shows the cusp-like minimum corresponds well to the obtained transition temperature. From these results, we can conclude that the field-induced magnetic ordering for $H \perp$ C-plane in IPACuCl$_3$ is described by the BEC of magnons. For $H \perp$ A-plane, as shown in Fig. \ref{Fig2} (a), the magnetic torque $\tau$ shows the minimum after decreasing and then becomes almost independent on temperature with the kink anomaly upon decreasing temperature. The arrows in Fig. \ref{Fig2} (a) also denote the transition temperature obtained from the phase diagram. We can see that the transition temperature obtained from the phase diagram does not correspond to the temperature at which the magnetic torque $\tau$ shows the minimum but to the temperature at which the magnetic torque $\tau$ shows the kink anomaly, differently from $H \perp$ C-plane. The same behavior has been observed in the temperature dependence of the magnetization for the magnetic field applied perpendicular to the easy-plane-type single-ion anisotropy vector $D$ in $S=1$ quasi-one-dimensional antiferromagnet NDMAP \cite{Honda}. For $H \perp$ B-plane as shown in Fig. \ref{Fig2} (b), it can be seen that the magnetization decreases and becomes almost independent on temperature upon decreasing temperature. Because the magnetic torque $\tau$ for $H \perp$ B-plane was measured down to $T=1.6$ K in the present experiments, the measurements at lower temperature are needed in order to investigate the behavior of the magnetic torque $\tau$ below $T_{\rm N}$ for $H \perp$ B-plane, as performed for $H \perp$ A- and C-planes. \par
As mentioned in Introduction, it is theoretically suggested that the Hamiltonian of systems is required to be rotationally invariant about the direction of the applied magnetic field to undergo the BEC of magnons \cite{Nikuni} so that the magnetic field should be applied parallel to the {\it easy-plane}-type single-ion anisotropy vector $D$, if exists, in order to observe the BEC of magnons in the field-induced magnetic ordered phase of spin gap systems. Because, there is the {\it easy-axis}-type fictitious single ion anisotropy $D^*$ perpendicular to the B-plane in IPACuCl$_3$, the BEC of magnons should not be observed in IPACuCl$_3$ from the theoretical suggestion. In order to understand the experimental observation of the BEC of magnons, we expect that the direction perpendicular to the A-plane is the second-easy-axis so that C-plane can be regarded as the {\it quasi-easy-plane}. In this situation, the BEC of magnons can be only observed in IPACuCl$_3$ when the magnetic field is applied perpendicular to the C-plane because the system retains rotational invariance, as observed in the present experiments. While, for $H \perp$ A-plane, this system has rotational invariance no longer so that this system does not undergo the BEC of magnons, and we can understand why we observed the same temperature dependence as that observed in NDMAP for the magnetic field applied in the easy-plane. Also, for $H \perp$ B-plane, we expect that the rotational invariance is broken so that the BEC of magnons is not observed. Actually, the rapid re-increase of the magnetic torque $\tau$ was not observed below $T_{\rm N}$ though the measurements were carried out down to $T=1.6$ K for $H \perp$ B-plane in the present experiments. To our best knowledge, this result, namely the BEC of magnons is only observed for a specific direction of applied magnetic field, is the first experimental evidence of breaking of the magnon BEC picture due to the rotational symmetry breaking. \par

\begin{figure}[t]
\begin{center}
\includegraphics[width=80mm]{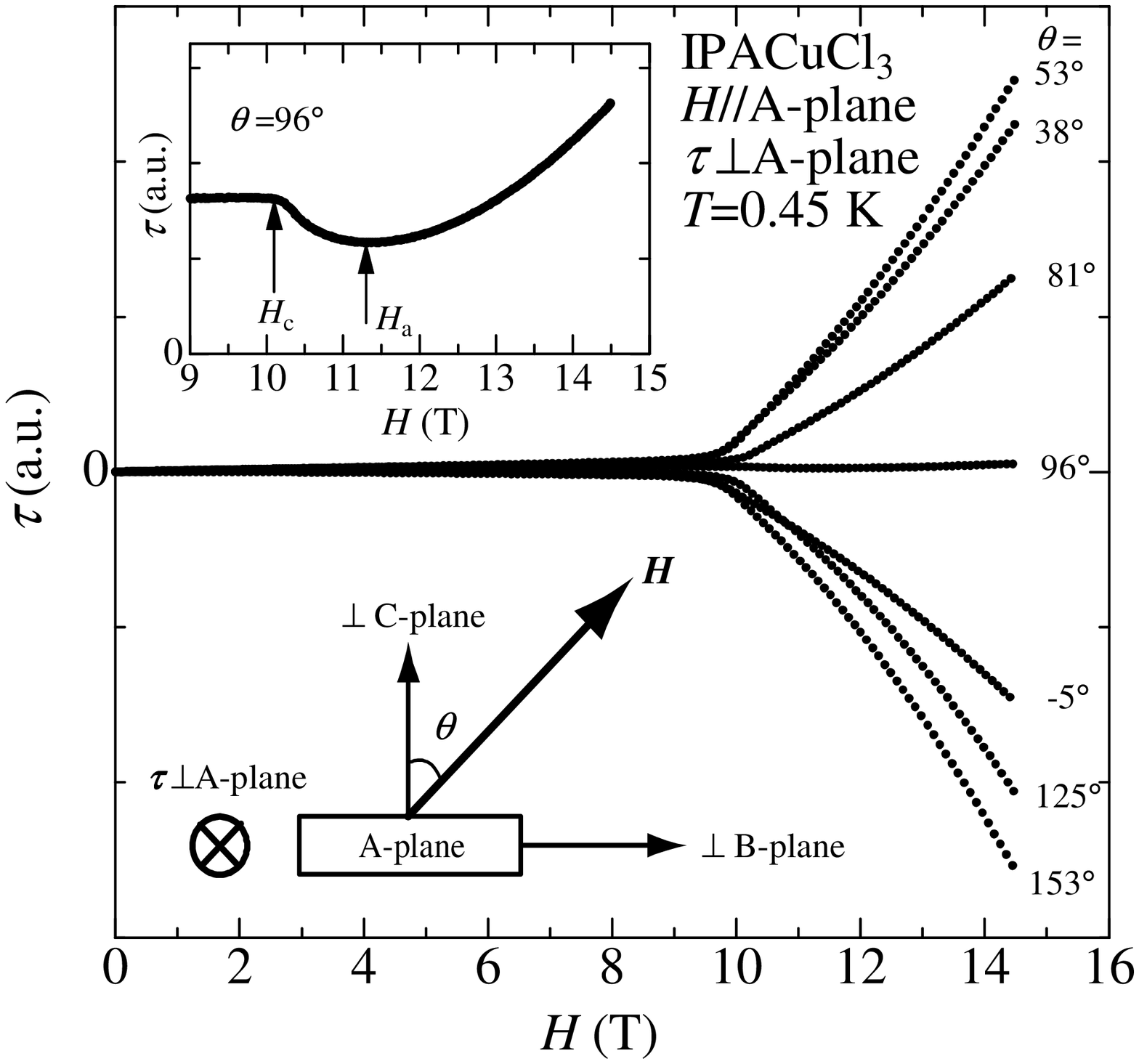}\\
\end{center}
\caption{Representative profiles of magnetic field dependence of the magnetic torque $\tau$ in IPACuCl$_3$ at various directions of the applied magnetic field in the A-plane. $\theta$ denotes the angle between the applied magnetic field and the direction perpendicular to the C-plane, as shown in the picture in this figure. The inset shows the enlarged profile of magnetic field dependence of the magnetic torque $\tau$ in IPACuCl$_3$ at $\theta=96^{\circ}$. \label{Fig3}}
\end{figure}    

Figure \ref{Fig3} shows the representative profiles of the magnetic field dependence of the magnetic torque $\tau$ in IPACuCl$_3$ at various angles of the applied magnetic field in the A-plane for $T=0.45$ K. As shown in Fig. \ref{Fig3}, the increasing of magnetic torque changes depending on the angle of the applied magnetic field, and the magnetic torque $\tau$ seems to retain almost zero at $\theta=96^{\circ}$, resulting that there is an easy-axis in the vicinity of this angle. Note that the angle is not the easy-axis exactly because the magnetic torque $\tau$ projected to the rotation axis of moving electrode, {\it i.e.} the direction perpendicular to the A-plane, can be only detected for each directions of applied magnetic field. By enlarging the data, the anomalous behavior clearly appears as shown in the inset of Fig. \ref{Fig3}, namely the magnetic torque $\tau$ shows the minimum at $H_a=11.3$ T and then changes to increase after the rapid decrease due to the closing of the spin gap at $H_c=10.1$ T with increasing magnetic field. Such behavior has been also observed at the magnetic field where the spin-flop transition occurs in magnetic torque measurements of other magnetic systems \cite{Sasaki2, Kajiyoshi} so that this may indicate the existence of the spin-flop transition in the magnetic field applied parallel to the easy-axis. The angle $\theta=96^{\circ}$ of the applied magnetic field is almost perpendicular to the B-plane, namely parallel to the easy-axis-type fictitious single ion anisotropy $D^*$ so that the spin-flop transition may occur when the ordered moments align parallel to $D^*$. The fact that the difference of magnitudes between $H_a$ and $H_c$ is comparable with the $D^* = - 0.24 \sim - 0.48$ T \cite{ManakaESR1, ManakaESR2} may also support that the anomaly at $H_a$ can be interpreted as the spin-flop transition. \par

\begin{fullfigure}[t]
\begin{minipage}{8.5cm}
\begin{center}
\includegraphics[width=80mm]{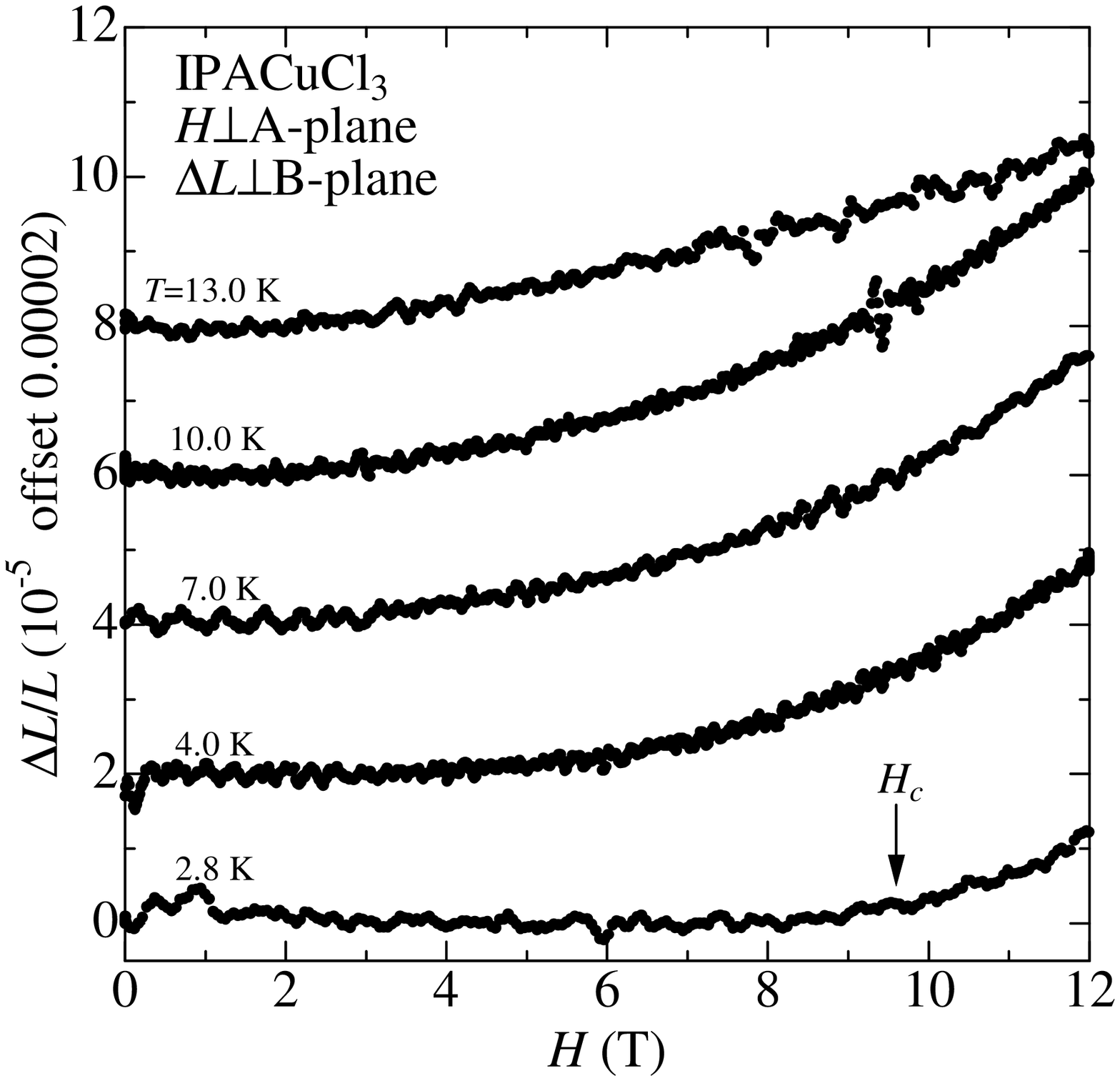}\\
(a)\\
\end{center}
\end{minipage}
\begin{minipage}{8.5cm}
\begin{center}
\includegraphics[width=80mm]{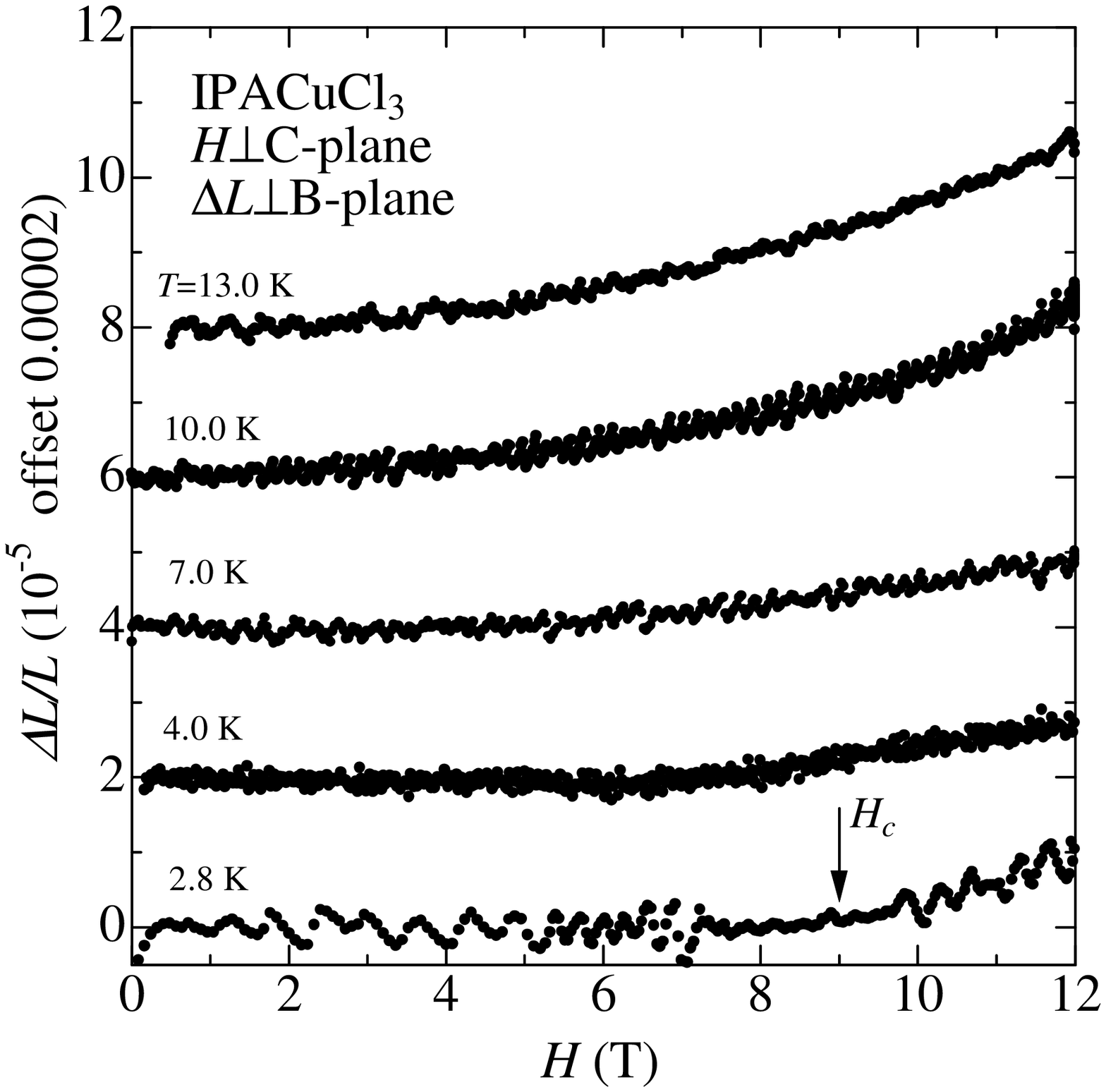}\\
(b)\\
\end{center}
\end{minipage}\\
\caption{Magnetostriction $\Delta L/L$ perpendicular to the B-plane for $H \perp$ (a) A-plane and (c) C-plane at various temperatures in IPACuCl$_3$. $H_c$ denotes the critical field corresponding to the spin gap \cite{Manakamag}. \label{Fig4}}
\end{fullfigure}

On the other hand, the additional phase transition in the field-induced ordered phase has been observed in NDMAP and discussed in terms of the spin-rotation transition or commensurate-incommensurate transition \cite{Tsujii}. Theoretically, Miyazaki {\it et al.} has demonstrated the spin-reorientation transition in the $S=1$ quasi-1D antiferromagnet with large easy-plane anisotropy when the magnetic field is applied at the intermediate angle, namely neither parallel nor perpendicular to the magnetic anisotropy axes \cite{Miyazaki}. It was interpreted that this transition is caused by the competition between the property of ordered moments that the field-induced staggered magnetization is already perpendicular to the applied magnetic field and the magnetic anisotropy, such as the single-ion anisotropy $D$ as well as the in-plane anisotropy $E$. The anomaly at $H_a$ was not observed in $H \perp$ B-plane though the measurements were carried out only above $T=1.6$ K, as shown in Fig. \ref{Fig1} (b) so that the anomaly at $H_a$ may indicate the spin-reorientation transition, rather than the spin-flop transition because the anomaly at $H_a$ was observed at the intermediate angle $\theta=96^{\circ}$, not $H \perp$ B-plane. However, the theory also predicts that the spin-reorientation transition occurs in a certain finite region of the angle and the magnitude of the critical field depends on the angle. In the present experiments, no anomaly at $H_a$ was observed at other angles though we measured at an interval of 15$^{\circ}$ of angle. Furthermore, no hysteresis indicative of the coexistence of ordered phases was observed though the theory predicts that the spin-reorientation transition is of the first-order. \par
The observed anomaly at $H_a$ in the present experiments have not been reported in the previous high-field experiments, such as the high-field magnetization processes \cite{Manakamag} and the specific heat measurements \cite{Manakaheat}. This may because the previous measurements were performed only in the applied magnetic fields perpendicular to A-, B- and C-plane. We expect that the anomaly will be also detected when the magnetic field is applied at the intermediate angles. \par      
As the other possibility, the consideration of the magnetostriction through the spin-lattice correlations may be also needed in order to understand the observed new anomaly. When the magnetostriction is induced by appearing of the magnetization due to the vanishing of the spin gap, the structural deformations induced by the magnetostriction may cause the changes of the fictitious single ion anisotropy $D^*$ and the anisotropy of the $g$-tensor so that the magnetic torque shows the variation or anomaly. In fact, such magnetostriction has been observed in the field-induced magnetic ordered phase of some spin gap systems \cite{Vyaselev,Sherman,Johannsen,Lorenzo,Sawai} and the importance of the spin-lattice correlations has been discussed. In order to investigate the existence, we performed the magnetostriction measurements for high magnetic fields in IPACuCl$_3$ and will discuss in the next subsection.

\subsection{Magnetostriction Measurements}

Figure \ref{Fig4} (a) and (b) show the magnetostriction $\Delta L/L$ perpendicular to the B-plane for $H \perp$ A- and C-planes at various temperatures in IPACuCl$_3$, respectively. As shown in Fig. \ref{Fig4}, the magnetostriction appears above the critical field $H_c$ and increases with increasing magnetic field, and the critical field $H_c$ decreases with increasing temperature. This behavior is similar to the field dependence of the magnetization \cite{Manakamag} and the magnetic torque, as shown in Fig. \ref{Fig1}, namely indicates that the magnetostriction is induced by emergence of the magnetization due to the vanishing of the spin gap through the spin-lattice correlations in IPACuCl$_3$. \par
\begin{figure}[t]
\begin{center}
\includegraphics[width=80mm]{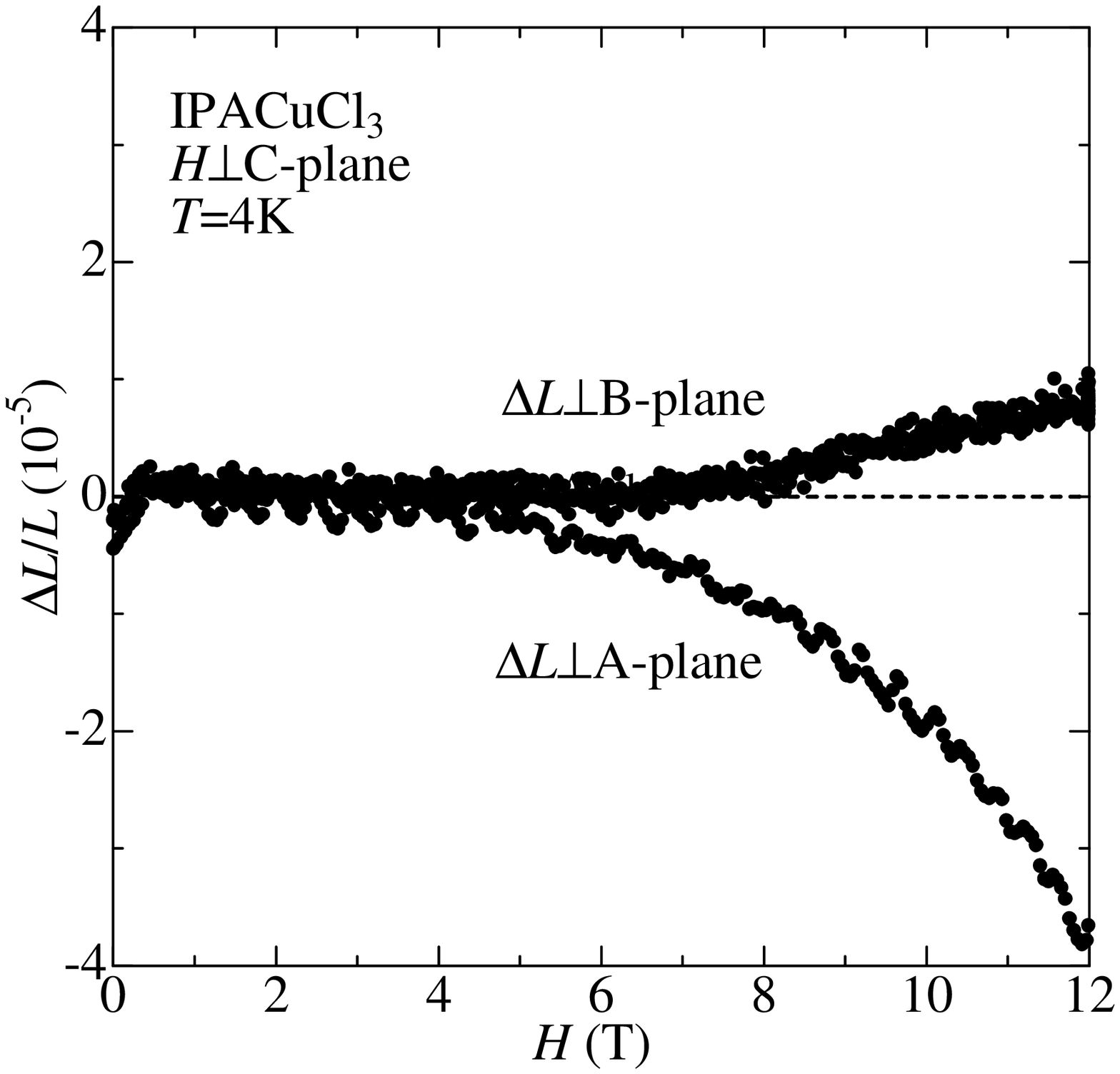}\\
\end{center}
\caption{Magnetostriction $\Delta L/L$ perpendicular to the A- and B-plane for $H \perp$ C-plane at $T = 4.0$ K in IPACuCl$_3$. \label{Fig5}}
\end{figure}

Figure \ref{Fig5} shows the magnetostriction $\Delta L/L$ perpendicular to the B- and A-plane as a function of the applied external magnetic field of $H \perp$ C-plane at $T = 4.0$ K in IPACuCl$_3$. As shown in Fig. \ref{Fig5}, $\Delta L/L$ perpendicular to the B-plane increases, while $\Delta L/L$ perpendicular to the A-plane decreases with increasing the magnetic field. The changed value of $|$$\Delta${\it L}/{\it L}$|$ in the direction perpendicular to the A-plane is four times as large as the case in the direction perpendicular to the B-plane. \par
Considering the spatial relation between the crystal grown surfaces (A-, B- and C-plane) and the crystal structure, {\it i.e.}, the $c$- and $b^*$-axes are perpendicular to the A- and C-planes, respectively, and the $a$-axis make an angle of $\sim 10^{\circ}$ in the C-plane with the direction perpendicular to the B-plane \cite{Manaka3DESR, ManakaMasuda}, the crystal lattice seems to expand to the direction of $a$-axis so that this distortion corresponds to the increase of the length between rungs of the ladder along the $a$-axis. Detailed diffraction measurements are needed in order to investigate and determine the variations of the crystal structure in high magnetic fields and we can expect that the spin-lattice correlations are discussed through the variations of the crystal structure in more details. 

\section{Conclusion}

We have presented the results of the high-field magnetic torque measurements in the spin gap system (CH$_3$)$_2$CHNH$_3$CuCl$_3$. The rapid increase of the magnetic torque $\tau$ above the critical field indicative of the field-induced magnetic transition due to the closing of the spin gap was clearly observed, as shown in Fig. \ref{Fig1}. From the temperature dependence of the magnetic torque $\tau$, the cusp-like minimum which is characteristic of the Bose-Einstein condensation of magnons was observed for $H \perp$ C-plane, while we observed the kink-anomaly for $H \perp$ A-plane, as shown in Fig. \ref{Fig2}. It was found that this different behavior between $H \perp$ A-plane and $H \perp$ C-plane can be understood by considering that the C-plane is regarded as the quasi-easy-plane and is the first experimental evidence of breaking of the magnon BEC picture due to the rotational symmetry breaking to our best knowledge. Also, the additional anomaly was observed in the field-induced magnetic ordered phase at the intermediate angle of the applied magnetic field in the A-plane, as shown in Fig. \ref{Fig3} and discussed in terms of the spin-flop transition, the spin-reorientation transition and the spin-lattice correlations. In fact, we performed the magnetostriction measurements and observed that the magnetostriction emerges rapidly by closing of the spin gap, which indicates the existence of the spin-lattice correlations in IPACuCl$_3$, as shown in Figs. \ref{Fig4} and \ref{Fig5}.     

\section*{Acknowledgements}

We acknowledge H. Manaka and T. Masuda for useful discussions and H. Oizumi for the help of the magnetic torque measurements. This work was supported by Grants-in-Aid for Scientific Research on Priority Areas "High Field Spin Science in 100 T" from the Ministry of Education, Science, Sports and Culture of Japan, the Saneyoshi Scholarship Foundation and the Kurata Memorial Hitachi Science and Technology Foundation. A part of this study was performed at the High Field Laboratory for Superconducting Materials, Institute for Materials Research, Tohoku University. \par

\end{document}